# LASER-INDUCED OPTICAL CHANGES IN AMORPHOUS MULTILAYERS

M. Malyovanik and S. Ivan

*Uzhgorod National University, Uzhgorod, Ukraine*

A. Csik

*Institute of Nuclear Research of the Hungarian Academy of Sciences, 4001 Debrecen P.O.Box 2., Hungary*

G.A. Langer and D.L. Beke[a)]

*Department of Solid State Physics, Debrecen University, 4010 Debrecen, P.O. Box 2., Hungary*

S. Kökényesi

*Department of Experimental Physics, Debrecen University, 4001 Debrecen, Hungary*

**Abstract**

It is shown that the well-known blue-shift of the fundamental absorption edge in as-deposited compositionally modulated amorphous Si/Ge and $As_6Se_{94}/Se_{80}Te_{20}$ multilayers (with periods of 4-8 nm) is further enhanced due to the thermal or laser-induced intermixing of adjacent layers. The laser-induced intermixing process, as supported by experiments and model calculations, can be attributed to both the local heating and photo-effects in $As_6Se_{94}/Se_{80}Te_{20}$ multilayers, while only the thermal effects were observed for Si/Ge multilayers. Structural transformations, based on this enhanced interdiffusion, provides good capability for spatially patterning optoelectronic devices and digital information recording.

[a)] Corresponding author.
*E-mail address*: dbeke@delfin.klte.hu

## Introduction

Technologies of nano-scale atomic engineering provide new possibilities in tailoring of properties of pure crystalline semiconductor materials (Si, Ge, GaAs), amorphous hydrogenated silicon (a-Si:H) or chalcogenide glasses as well.[1-5] Besides the great amount of investigations on nanocomposites, compositionally modulated multilayers are attractive because of the well-developed microelectronic technology, rather simple modeling of physical processes and possible applications of thin film structures. Although the misfit problems in amorphous multilayers (AML) are considerably reduced as compared to crystalline superlattices, still there are some physical processes (e.g. quantum confinement, diffusion) not well understood.[4-6]

The structure, optical and electrical characteristics of amorphous Si and Ge layers are widely studied,[7] and e.g. our previous investigations on laser-induced structural transformations of amorphous Si/Ge multilayers resulted an interesting results.[8] It was shown that these AMLs were stable against of crystallization, but undergo other structural transformations under high power laser irradiation ($\lambda$ = 0.63 μm, $P$ = 1-100 W/cm$^2$) at room temperature which can be attributed to the local heating. Furthermore, from a comparison with the behavior of amorphous Se/$As_2S_3$ multilayers, it was concluded that – in contrast to the Si/Ge system – in this case the photo-stimulated interdiffusion (without direct heating) played an important role in the change of the optical properties, i.e. such a comparison can bring additional information about the role of photo-induced structural effects in intermixing as well.[5, 8, 9] The photo-stimulated interdiffusion in a-Se/$As_2S_3$ system[8, 9] can be used for surface hologram recording. On the other hand there are no publications about similar process in a-Si/Ge, although thermo-stimulated interdiffusion can lead to structural changes in these multilayers as well.[8,11,12]

Diffusion processes at interfaces in general are also related to the problem of thermal stability of heterogeneous systems, which is very important in nanostructures, especially under additional external influences (light, fields, heat flow).[8-11] It is also known, that amorphous solid solutions can be produced by diffusional intermixing of AML of wide (Si) and narrow (Ge) bandgap semiconductors (barriers and valleys, respectively)[13,14] and there is a similar situation for the As-Se-Te system of chalcogenide glasses.[15] In the latter case amorphous $As_6Se_{94}$ (transparent barrier, stable to crystallization) and $Se_xTe_{1-x}$ (absorptive, narrow-band valley) can also form solid solutions during intermixing.

The aim of this work is to extend our preliminary results on a-Se/$As_2S_3$ and a-Si/Ge system[8], and to determine the characteristics and mechanism of the light-or heat-induced structural changes and interdiffusion as well as their interrelation with the optical parameters in a-Si/Ge and $As_6Se_{94}/Se_{80}Te_{20}$ systems. Possible applications for optical information recording for these results are also considered.

**Experimental**

Si/Ge multilayers with a total thickness up to 0.1-0.3 μm and modulation periods of $\Lambda$ = 4-6 nm were prepared by DC magnetron sputtering onto sapphire or quartz substrata.[8] $As_6Se_{94}/Se_xTe_{1-x}$ AML ($0.6 < x < 0.9$), with total thickness up to 0.8-2.0 μm and modulation period of $\Lambda$ = 5-8 nm, were deposited by cyclic thermal evaporation in vacuum onto the same type of substrata.[16] Besides Si/Ge the chalcogenide system with $Se_{80}Te_{20}$ "well" layers were selected as most suitable for measurements at the given conditions (time domain of expositions; measurements at optimal absorption of He-Ne laser light with $\lambda$ = 0.63 μm; stability to crystallization). Small Angle X-Ray Diffraction (SAXRD) was used to investigate the periodicity and the quality of interfaces. Few cross-sections were made and investigated by TEM (JEOL 2000) and also proved that the applied technology is capable to produce good multilayers (Fig. 1).

The SAXRD method was also used for the investigations of interdiffusion.[6,8,16] The change of the optical transmission, $\tau$, (focused He-Ne laser beam of 0.2 mm in diameter, $P$ = 0.03-30 W/cm$^2$ density of surface intensity and $\lambda$ = 0.63 μm) with illumination time $t$ was measured *in situ* in a special cryostat at temperatures between 100 K and 300 K. Ar-ion laser was also used for illuminate the chalcogenide samples. Measurements of $\tau$, at the same $\lambda$ but with a small intensity beam, were also carried out on samples previously annealed at different fixed temperatures between 350-500 K, and the results were compared with those of the SAXRD measurements. The optical transmission spectra obtained in as-deposited and heat-treated AML samples were compared with transmission spectra of homogeneous Si, Ge, $As_6Se_{94}$, $Se_{80}Te_{20}$ layers, having the same total thickness as the corresponding multilayers.

**Results and discussion**

Blue-shift of optical absorption edge was observed in all as-deposited multilayers. In the analysis of optical transmission data the "effective optical medium" model[17] (in which mainly the "well" layers, with small $E_{gv}$ energy, determine the optical absorption $\alpha = 10^2$-$10^5$ $cm^{-1}$ at the average absorption edge $E_g$, and the "barrier" layers with larger $E_{gb}$ are transparent) has been applied. Since in this model the absorption is mainly determined by the "well" layers, the relation

$$\alpha h\nu = const. \cdot \left(h\nu - E_g\right)^2 \quad (1)$$

can be used for the determination of $E_g$ in amorphous semiconductors, where $h\nu$ is the energy of light quantum.[7] Similarly to other AMLs[3,4] the observed $\Delta E_g$ = 0.13-0.16 eV blue-shift (the reference state was a thick homogeneous layer, having the same composition as the "well" layers) in our as-deposited multilayers (see Fig. 2a for illustration), can be connected with quantum-confinement effects.

Furthermore in accordance with the same model [17] with decreasing thickness of the "well" layers, and/or with the appearance and growth of intermixed layers due to interdiffusion, further $\Delta E_g$ shift (0.05 ÷ 0.1 eV) and further bleaching of the AML was observed (Fig. 2b). At the same time the optical reflection coefficient $R$ of our effective optical medium decreases, because the resulting mixture has smaller density and refraction index.

Photo-induced changes of $\tau$ during laser illumination at $\lambda = 0.63$ μm with $P < 1W \cdot cm^{-2}$ were not observed in homogeneous a-Si, a-Ge thin films between 150 – 400 K. The $Se_{80}Te_{20}$ layers also were not light-sensitive, which is characteristic for Te-containing chalcogenide layers.[18] Small photo-darkening ($\tau/\tau_o \leq 10\%$) was observed at 293 K in $As_6Se_{94}$ individual layers, which is in accordance with the known dependence of the sensitivity on the composition and temperature in the As-Se system.[18, 19] This also indicates that photo-darkening may decrease with increasing temperature in $As_6Se_{94}/Se_{80}Te_{20}$ multilayers due to the increasing efficiency of the thermal erasement of photo-induced structural changes in $As_6Se_{94}$ sub-layers. Indeed, photo-darkening was not observed above 300 K in our chalcogenide AMLs. However cooling down the multilayers photo-darkening appears and even can prevail, at any rate, in the initial stage of the exposition in $As_6Se_{94}/Se_{80}Te_{20}$ system. As it can be seen on the curves 1-4 of Fig. 3 photo-induced structural changes in $As_6Se_{94}$ sub-layers are faster and more efficient (decreasing part of the curve) than those related to the intermixing (increasing part). The resulting degree of photo-darkening is less expressed than the photo-bleaching caused by further long-term intermixing except of very low temperatures, where the diffusion is extremely slow. In a-Si/Ge system the bleaching disappears below 170 K even at the maximum value of $P$ used (see the 1' curve in Fig. 3), but photo-darkening does not appear. It means, that photo-induced structural changes are not essential in Si/Ge AMLs, in spite of the fact, that electron-hole pairs are also generated by illumination in these semiconductors.

The following explanation of laser-induced structural transformations in AMLs can interpret our results. It is plausible to assume that both thermally activated as well as photo-induced structural changes contribute to the process. *Photo-induced structural changes* are inherent in chalcogenide structures and are caused both *by the usual photo-induced changes* (already observed in a homogeneous, light-sensitive amorphous chalcogenide phase), and *by the stimulated interdiffusion* between the adjacent layers. The light-stimulated increase of the rate of intermixing is obviously related to the light-induced viscous flow in chalcogenide glasses [20] and is proportional to the number of excited chalcogen-bridge atoms, which determine the local deformations due to the bond switching and displacements.[18, 19] Note that besides the increase of the number of the excited defects the change of the migration activation energy $E_a$ can also influence the diffusion coefficient, *D,* but the mechanism of this in illuminated chalcogenide glasses is not well established at present. During normal or laser heating the temperature of the samples increases and, since the *D* increases with increasing temperature, the *thermally-stimulated interdiffusion is a common part of laser-induced structural transformations* in all AMLs, where intermixing by the formation of solid solutions takes place. Our results support this, which can also explain the deviations from the linearity of optical response of the sensitive medium on the light intensity as follows.

In chalcogenide-based AMLs both heat- and light stimulated interdiffusion processes must be taken into consideration, but when the temperature rises above 300 K the photodarkening disappears, increasing additionally the bleaching. The contribution of the thermally induced intermixing can be measured by the deviation from the linearity (dotted line in Fig. 4) of the *E(P)* function (*E* - exposition, *P* - intensity) taken for a fixed photobleaching level (Fig. 4, curves 1, 2). In a-Si/Ge multilayers the intermixing process can be observed in a real time scale only at $P \geq 1$ W/cm$^2$, where thermal processes determine the interdiffusion.

The temperature in the illuminated spot can be determined from the comparison of the $\tau/\tau_0 = f(t)$ curves, obtained at different temperatures (provided by the external heating of the samples) and at low-intensity illumination, with the $\tau/\tau_0 = f(t)$ curves, measured at room temperature but at different illumination intensities (triangles 3, 4 in Fig. 4). The temperatures determined above are in good agreement with temperatures, calculated using the known equations for the temperature of the layer surface illuminated by an intense laser beam,[21,22] as follows (see continuous curves 3, 4 in Fig. 4). For the Gaussian intensity distribution inside the laser beam with a radius $r$ ($P = P_0 \cdot exp\{-r^2/w^2\}$, $w$ is the effective radius of the beam, $P_0$ is the intensity at the center of the surface spot), one can calculate the maximum temperature $T_{\max}$ in the thin, slightly absorbing film. If $l << w$, $k_s w >> k_f l$ ($l$ - thickness of the layer, $k_s$ and $k_f$ are heat conductivity's of the substratum and the layer):

$$T_{\max} = \frac{P_a}{2\sqrt{\pi} w k_s}. \tag{2}$$

Here $P_a = P_0 \alpha l \cdot (1-R)$ is the absorbed intensity, $R$ – the reflectivity of the surface. Equation (2) is applicable for our AML made of chalcogenide glasses, since here the inequality $k_s w >> k_f l$ fulfills.

On the other hand if $l << w$ and $k_s << k_f$ (which is valid for Si/Ge multilayers), then

$$T_{\max} = \frac{P_a}{2\sqrt{\pi} l k_f} \int_0^{\infty} J_0\left(\lambda, {r}/{w}\right) \cdot \exp\left(-0.25 \cdot \lambda^2\right) \cdot \left( \frac{k_f}{k_s} \cdot \frac{1}{\lambda\left(\frac{k_f}{k_s} - 1\right) - \frac{1}{l}} \right) d\lambda, \tag{3}$$

where $J_0$ is the Bessel function of order zero. For Si/Ge AML the integral in Eq. (3) equals approximately to 1, so the resulting temperature $T_{max}$ at the center of the spot can be expressed as $T_{max} = P_a/2\pi l k_f$. The differences between $T_{max}$ calculated according to Eqs. (2) and (3) (curves 3, 4 in Fig. 4), and from the data on $\tau/\tau_0 = f(t)$ plots, connected to the interdiffusion process (triangles in Fig. 4), may be attributed to the additional influence of strains on interdiffusion [6] and to the deviation of calculated $T_{max}$ from the temperature obtained by averaging over the temperature distribution in the spot (since the optical transmission is not fully uniform, according to the distribution of the laser-beam intensity and the temperature in the spot). It can also be seen in Fig. 4 that the deviation between the calculated and experimental curves is larger for $As_6Se_{94}/Se_{80}Te_{20}$ then for Si/Ge multilayers, especially at high laser powers. However, even the deviation shown for the chalcogenide system is still below or very close to the experimental errors of the points (see the error bars in Fig. 4).

The measured 10-20% changes of optical transmission in investigated AMLs are accompanied by the corresponding changes of optical reflection coefficient $R_{op}$ ($\Delta R_{op}/R_{op} \approx 20\%$), refraction index and even of the total thickness of AML up to 1-3%. All these can be used for optical data recording, creation of phase modulation structures, surface reliefs by localized laser irradiation. Optimization of AML nanostructures for optical recording includes the selection of combined pairs of materials, creating solid state solutions during interdiffusion, enhanced by light and/or heating. The nanometer-range of compositional modulation must be accounted to ensure maximum sensitivity at illumination and stability of AML in the dark (at 293 K). The efficiency of the amplitude modulation ($\Delta\alpha$, $\Delta R$), is comparable with recording parameters in the photo-crystallizing chalcogenide layers, but the phase modulation due to the change of the refractive index and thickness[8, 9] within amorphous state of multilayer broadens the application possibilities. All these enables creation of optical or geometrical reliefs in the AML, which can be used in optoelectronics as diffractive elements, microlenses, optical waveguides, memory elements

**Summary**

Amorphous Si/Ge and $As_6Se_{94}/Se_{80}Te_{20}$ multilayers with compositional modulation periods of 4-8 nm possess effect of blueshift of the fundamental absorption edge in as-deposited structures, which can be further enhanced due to the thermal or laser-induced intermixing of adjacent layers. Two components of the induced intermixing process in chalcogenide AML were distinguished and attributed to the local heating and photo-effects in $As_6Se_{94}/Se_{80}Te_{20}$ multilayers, but only the thermal effects were identified for Si/Ge samples. This latter behavior is universal for all multilayers where solid solutions may be created. Structural transformations based on the enhanced interdiffusion in these nanostructures provides the capability of spatially patterning optoelectronic devices, optical information recording.

**Acknowledgment**

This work was supported by OTKA Grants No. T-025261, T-038125 and T-037509, by the grant of the Hungarian Ministry of Education FKFP No. 0325/2000. Authors would like to thank Shipljak M. for his assistance in preparation of samples.

## References

[1] W. K. Choi, W. Ng, V. S. L. Swee, C. S. Ong, M. B.Yu, and R. S. F. Yoon, Scripta Mater. **44**, 1875 (2001)

[2] K. Brunner, G. Abstreiter, M. Walther, G. Bohm, and G. Tranke, Surface Sci. **267**, 218 (1992)

[3] M. Hirose and S. Miyazaki, *Japan Annual Reviews in Electronics, Computers and Telecommunications*, ed. Y.Hamakawa, OHM-North-Holland **22**, 147 (1987)

[4] D. Nesheva, D. Arsova, and Z. Levi, Phil.Mag. **B69**, 205 (1994)

[5] A. Kikineshi and A. Mishak, in *„Physics and Applications of Non-Crystalline Semiconductors in Optoelectronics"* NATO ASI Series **36**, 249 (1997)

[6] A. L. Greer, Defect and Diffusion Forum **129-130**, 163(1996)

[7] N. F. Mott and E. A. Davis, *Electron Processes in Non-Crystaline Materials*, Clarendon press, Oxford (1979)

[8] A. Csik, M. Malyovanik, J. Dorogovics, A. Kikineshi, D. L. Beke, I. A. Szabo, and G. Langer, J. Optoelectronics and Advanced Materials **3**, 33 (2001)

[9] V. Palyok, A. Kikineshi, I. Szabo, and D. L. Beke, Appl.Phys. **A 68,** 489 (1999)

[10] D. L. Beke, G. A. Langer, A. Csik, Z. Erdélyi, M. Kis-Varga, I. A. Szabó, and Z. M. Papp, Defect and Diffusion Forum **194,** 1403 (2001)

[11] A. Csik, G. A. Langer, D. L. Beke, Z. Erdelyi, M. Menyhard, and A. Sulyok, J. Appl. Phys. **89,** 804 (2001)

[12] D. L. Beke, A. Dudas, A. Csik, G. Langer, M. Kis-Varga, L. Daroczi, and Z. Erdelyi, Functional Materials **6**, 539 (1999)

[13] T. S. Massalsky (ed. in chief), *Binary alloy phase diagrams*, Second ed. ASM International, 1992, p. 2000

[14] S. M. Prokes and F. Spaepen, Appl. Phys. Lett. **47**, 234 (1985)

[15] Z. U. Borisova, *Chalcogenide Semiconductor Glasses*, LDU Publ., Leningrad (1983)

[16] A. Imre, V. Fedor, M. Kis-Varga, A. Mishak, and M. Shiplyak, Vacuum **50**, 507 (1998)

[17] P. Berning, in *„Physics of Thin Films"*, ed. H. Hasse, Moscow, Mir (1967)

[18] A. Kikineshi, Optical Memory and Neural Networks **4**, 177 (1995)

[19] A. Kolobov and K. Tanaka, J. Optoelectronics and Advanced Materials **1**, 3 (1999)

[20] D. K. Tagantsev and S. V. Nemilov, Fiz.Khim.Stekla **15**, 397 (1989)

[21] E. Abraham and J. M. Halley, Appl. Phys. A. **42**, 279 (1987)

[22] M. J. Lax, Appl. Phys. **48**, 3919 (1977)

**Figure captions:**

FIG 1. TEM picture (a) and small angle X-Ray specrum (b) of  as-deposited Si/Ge multilayers.

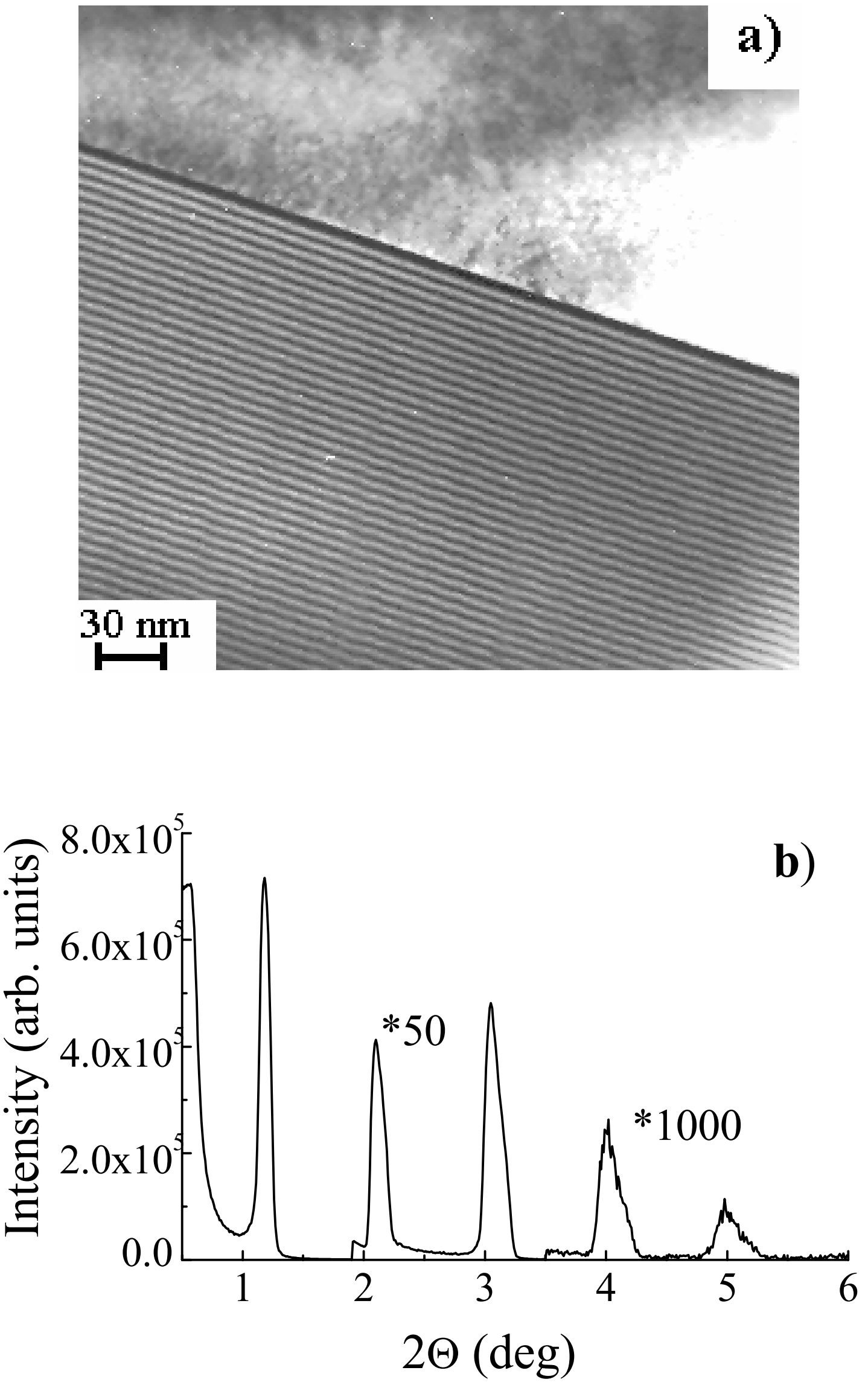

FIG 2a. Optical absorption spectra of a-Ge (1) and $Se_{80}Te_{20}$ (3) thin homogeneous layers and Si/Ge AML before (2) and after (2') 6 hours illumination with Ar-ion laser at 0.51 μm, $P$ = 7 W/cm$^2$ and for $As_6Se_{94}/Se_{80}Te_{20}$ AML before (4) and after (4') 30 minutes illumination with He-Ne laser at 0.63μm, $P$ = 0.2 W/cm$^2$.

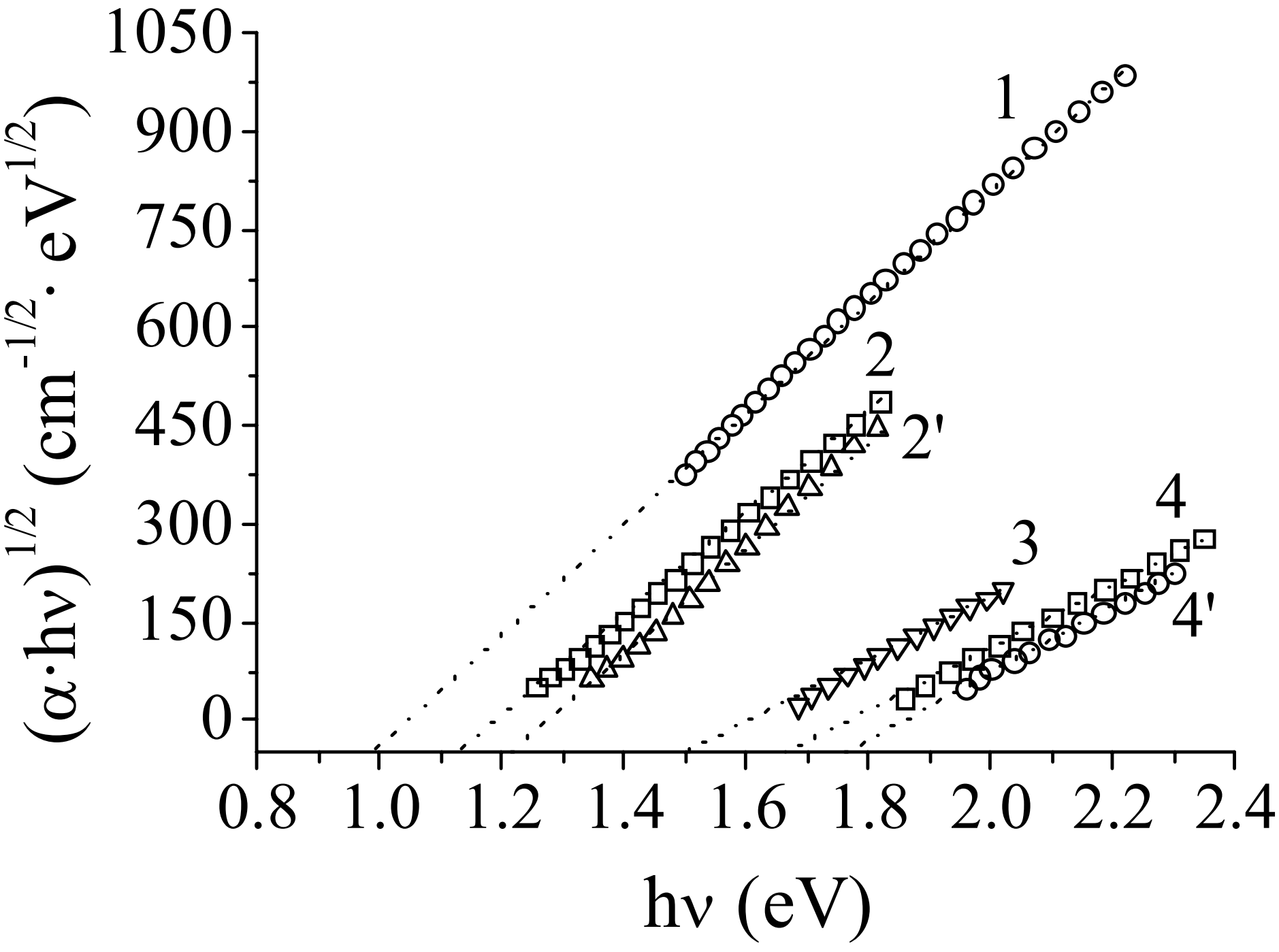

FIG 2b. Optical gap dependence on the modulation period for Si/Ge (1) and $As_6Se_{94}/Se_{80}Te_{20}$ (2) AML.

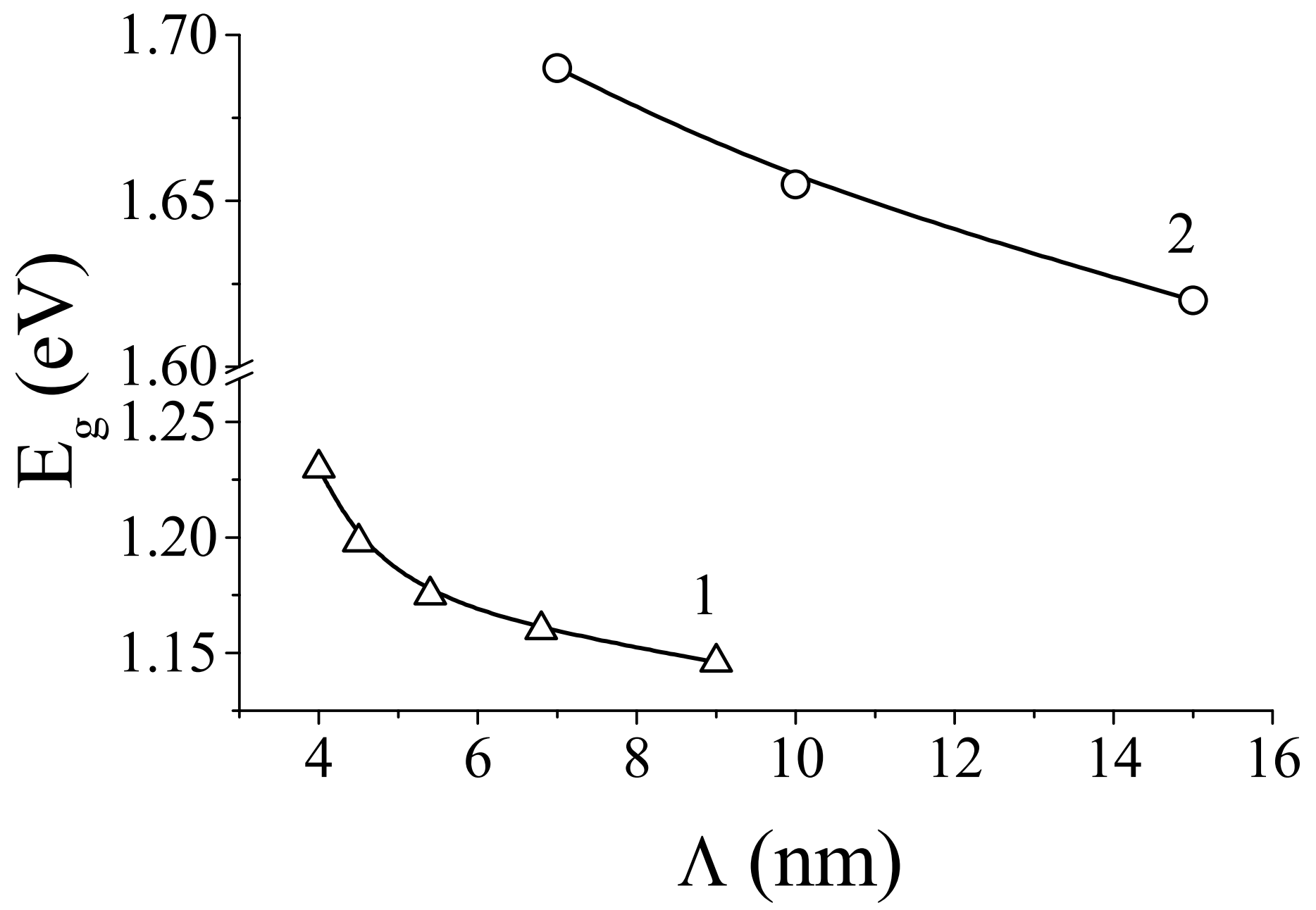

FIG 3. Relative change of the optical transmission for Si/Ge and $As_6Se_{94}/Se_{80}Te_{20}$ AML at T = 293 K (2') and 170 K (1') as well as at 100 K (1), 170 K (2), 210 K (3) and 293 K (4), respectively during illumination with focused He-Ne laser light ($P$ = 28 W/cm$^2$, $\lambda$ = 0.63 μm). The upper time-scale is for $As_6Se_{94}/Se_{80}Te_{20}$ AML, the lower for Si/Ge.

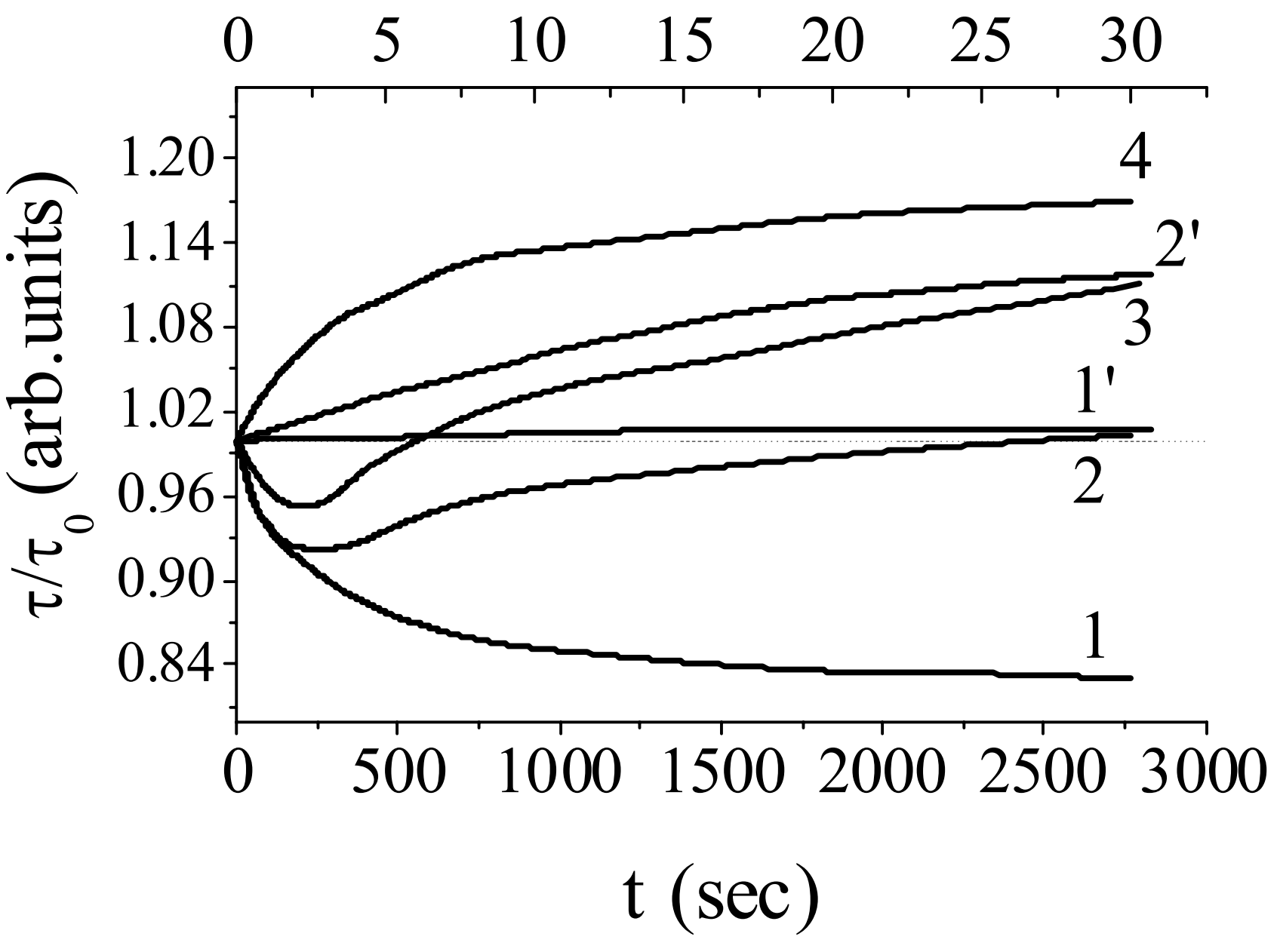

FIG 4. The change of exposition *E* for equal photobleaching level at 293 K upon illumination intensity for Si/Ge (1) and $As_6Se_{94}/Se_{80}Te_{20}$ (2) AML. The dependence of the maximum temperature in the center of the illuminated spot on the surface of Si/Ge (3) and $As_6Se_{94}/Se_{80}Te_{20}$ (4) AML upon the illumination intensity.

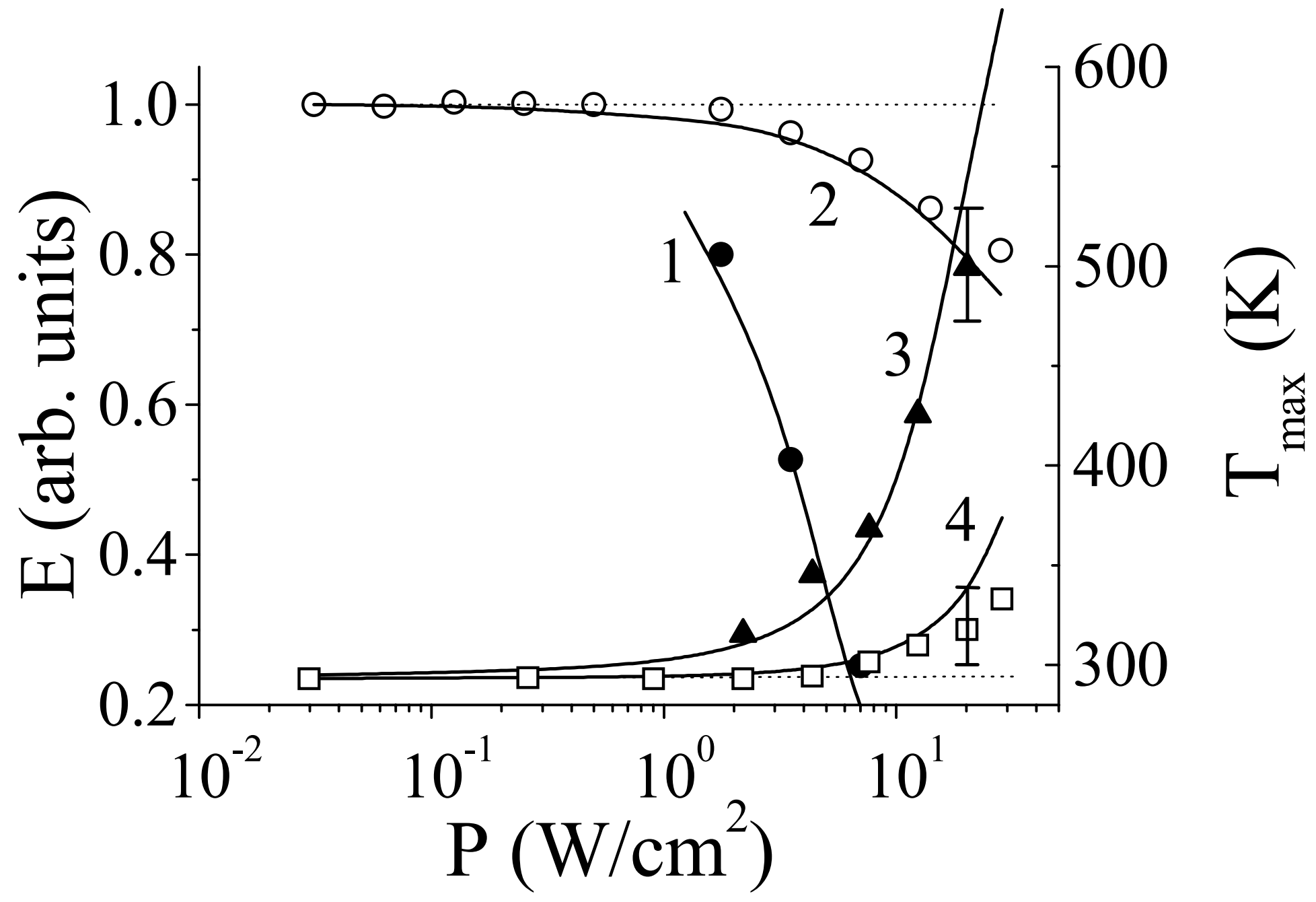